\documentclass[preprint,aps]{revtex4} 

\usepackage{epsf}
 
\begin{document}

\title
{
High-Order Coupled Cluster Method Calculations using Three-Dimensional Model 
States: An Illustration for the Triangular-Lattice Antiferromagnet in an External Field
}
\author
{
D. J. J. Farnell and A. I. Croudace
}
\affiliation
{
Division of Mathematics and Statistics, Faculty of Computing, Engineering, and Science, 
University of South Wales, Pontypridd CF37 1DL, Wales, United Kingdom
}


\begin{abstract}
The coupled cluster method (CCM) has previously been applied to 
study the ground- and excited-state properties of many different 
types of frustrated and unfrustrated quantum spin systems. A common
feature in the application of the CCM is to rotate the local spin axes of the 
(often classical) model state so that (notationally only) the spins all appear 
to point in the downwards $z$-direction. Hitherto, 
we remark that only coplanar model states have been used because they 
do not lead to imaginary terms in the new Hamiltonian. By contrast, non-coplanar 
``three-dimensional'' (3D) model states can lead to imaginary terms in the new 
Hamiltonian after rotation. In principle, however, macroscopic quantities 
predicted by the CCM (such as the ground-state energy and order 
parameter) should still be real (even though the Hamiltonian may be complex)
because the transformations of local spin axes are unitary. Here we 
explain how we may use such 3D model states for the CCM 
and how we may solve for the (now possibly complex) CCM correlation 
coefficients. We present results for the spin-half triangular-lattice antiferromagnet 
in an external magnetic field. We use both coplanar model states and a single 
3D non-coplanar model state. We calculate the ground-state 
energy and the lattice magnetization as a function of the magnetic field 
strength. We find that the energies
of the coplanar states lie lower than that of the non-coplanar state for
all values of the external field, as seen in other approximate studies for this model. 
We find that  the non-coplanar state does not detect the well-known spin plateau 
occurring in this model, although (as seen before) it is clearly observed 
for the coplanar model states using the CCM. These results are an
excellent initial validation of the new approach for the application of the CCM 
using 3D model states.
\end{abstract}


\maketitle

\section{Introduction}

In this article we apply  a method of quantum many-body theory called 
the coupled cluster method (CCM)
\cite{ccm1,ccm2,ccm5,ccm12,ccm15,ccm20,ccm26,ccm27,ccm32,ccm35} 
to study strongly interacting quantum spin systems.  The CCM is not 
restricted, in principle, by the spatial dimensionality of the
problem or by the presence of competition between bonds, i.e., in 
frustrated quantum spin systems. An important  advance in the accuracy 
of the method for a localized approximation scheme called the LSUB$m$ 
scheme has been afforded by the use of ``high-order'' CCM via
computer-algebraic implementations \cite{ccm12,ccm15,ccm20,ccm26}. 
This computer code developed by DJJ Farnell and J Schulenburg \cite{code} 
is very flexible in terms of the range of underlying 
crystallographic lattices, values for spin quantum number, and types
of Hamiltonian that may be studied. A common first task
in the practical application of the CCM is to rotate the local spin axes of the 
(often classical) model state so that (notationally only) the spins all appear 
to point in the downwards $z$-direction.  Although the Hamiltonian is changed
by transforming these local spin axes, these rotations are unitary and 
so they do not affect the energy eigenvalues or expectation values. 
Furthermore, we note that hitherto only coplanar model states have been used 
because they do not lead to complex numbers in the new Hamiltonian.
By contrast, three dimensional (3D) non-coplanar model states can 
lead to imaginary terms in the new Hamiltonian after rotation and so 
are more difficult to treat computationally. In principle, however, macroscopic 
quantities predicted by the CCM (such as the ground-state energy and order 
parameter) should be still real even though the Hamiltonian is now complex 
because again these transformations of local spin axes are unitary. Here we 
explain how we may use such 3D model states for the CCM 
and how we may solve for the (now possibly complex) CCM correlation 
coefficients. 

As an illustration of the method, we consider the spin-half triangular-lattice
Heisenberg antiferromagnet in a magnetic field \cite{LhuiMi,hon1999,CGHP,squareTriangleED,HSR04}. 
As is well-known, the response of quantum magnetic systems to an external 
field is revealed by its magnetization curve. The magnetic processes of
quantum anitferromagnets is discussed, e.g., in Refs. \cite{hon1999,squareTriangleED,HSR04,nishi,chub,alicea,oshi,SchuRi,jump,ono,kagome_pl,schnalle,schroeder,fortune}, and the interested reader is referred to these sources for more 
details. The Hamiltonian that we will use here is given by 
\begin{equation}
H = \sum_{\langle i,j\rangle} {\bf s}_i ~ \cdot ~ {\bf s}_j - \lambda \sum_i s_i^z
~~ ,
\label{heisenberg}
\end{equation}
where the index $i$ runs over all lattice sites on the triangular lattice. The
expression $\langle i,j\rangle$ indicates a sum over all nearest-neighbor 
pairs, although each pair is counted once and once only. The strength 
of the applied external magnetic field is given by $\lambda$.

There are two ground states classically (shown in Fig. \ref{model_states}): 
a set of coplanar states and a single non-coplanar state. The quantum system is discussed in Refs. 
\cite{hon1999,HSR04,nishi,chub,alicea,chub94,trumper00,ono,squareTriangleED}. 
Although on the classical level both cases (coplanar and 
non-coplanar) are energetically
equivalent\cite{kawamura,chub,zhito,cabra}, 
previous results of approximate methods indicate that 
thermal or quantum fluctuations ought to favor the planar configuration
\cite{kawamura,chub,zhito,cabra}. 
Previous results of the CCM \cite{farnell} indicate that a plateau state 
occurs for  $1.37 \lesssim \lambda \lesssim 2.15$. (Note that we compare
new results presented in this article for the non-coplanar states to those
earlier results of Ref. \cite{farnell}.) These results for the plateau 
are in excellent agreement with experimental
results for the magnetic compound Ba$_3$CoSb$_2$O$_9$ 
(a spin-half triangular-lattice antiferromagnet), 
which demonstrates a spin plateau that agrees quantitatively with
results of the CCM and exact diagonalizations \cite{shirata}.
Furthermore, CCM results indicate that a similar plateau occurs over 
the range $2.82  \lesssim  \lambda  \lesssim  3.70$ for the for the spin-one 
triangular-lattice antiferromagnet, and this theoretical result has subsequently 
been established experiment for the compound Ba$_3$NiSb$_2$O$_9$ 
(a spin-one triangular-lattice antiferromagnet) \cite{richter}.

The main goal of our paper is to explain how the CCM can be used with
3D model states. Firstly we present a brief description of the CCM  formalism and its 
application via computational methods to the subject of quantum spin models with
3D model states. As an illustration of our method, we describe the application of 
the method to the spin-half Heisenberg model for the triangular lattice at zero 
temperature in the presence of an external magnetic field. We present our 
results and then discuss the conclusions of this research.

\section{The Coupled Cluster Method (CCM)}

\begin{figure}
\epsfxsize=12cm
\centerline{\epsffile{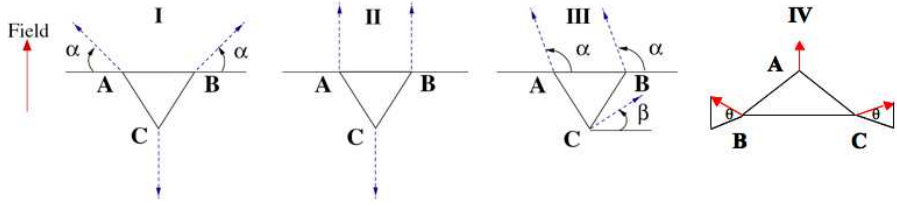}}
\caption{Classical ground states (also the CCM model states): 
I to III are coplanar, whereas IV is non-coplanar with spins 
at an angle $\theta$ to the plane perpendicular to the external field.}
\label{model_states}
\end{figure}

As the CCM has been discussed extensively elsewhere (see Refs. 
\cite{ccm1,ccm2,ccm5,ccm12,ccm15,ccm20,ccm26,ccm27,ccm32,ccm35}),  
a brief overview of the method is presented here only. Note however that 
the solution of the CCM equations for the case of 3D model states
is presented in an Appendix, which has not been attempted before. 
In this case, CCM correlation coefficients may be complex, and so 
extensive changes to the basic computer code that implements the
CCM to high orders of approximation are necessary, again as described 
in the Appendix.
We begin the brief overview of the CCM method by presenting the 
ground-state Schr\"odinger equations, which are given by
\begin{equation} 
H |\Psi\rangle = E_g |\Psi\rangle
\;; 
\;\;\;  
\langle\tilde{\Psi}| H = E_g \langle\tilde{\Psi}| 
\;. 
\label{eq1} 
\end{equation} 
and bra and ket states are given by
\begin{eqnarray} 
|\Psi\rangle = {\rm e}^S |\Phi\rangle \; &;&  
\;\;\; S=\sum_{I \neq 0} {\cal S}_I C_I^{+}  \nonumber \; , \\ 
\langle\tilde{\Psi}| = \langle\Phi| \tilde{S} {\rm e}^{-S} \; &;& 
\;\;\; \tilde{S} =1 + \sum_{I \neq 0} \tilde{{\cal S}}_I C_I^{-} \; .  
\label{eq2} 
\end{eqnarray} 
We use model states (denoted $|\Phi\rangle$ in the 
ket state and $\langle\Phi|$ in the bra state) 
as references states for the CCM and those 
used here are shown in Fig. \ref{model_states}.

We note that the CCM ket- and bra-state equations are given by
\begin{eqnarray} 
\langle\Phi|C_I^{-} {\rm e}^{-S} H {\rm e}^S|\Phi\rangle &=& 0 ,  \;\; 
\forall I \neq 0 \;\; ; \label{ket_state_eqn} \\ 
\langle\Phi|\tilde{S} {\rm e}^{-S} [H,C_I^{+}] {\rm e}^S|\Phi\rangle 
&=& 0 , \;\; \forall I \neq 0 \;\; . \label{bra_state_eqn}
\end{eqnarray} 
and that the method in which Eqs. (\ref{ket_state_eqn}) 
and (\ref{bra_state_eqn}) are solved has been discussed extensively 
elsewhere \cite{ccm1,ccm2,ccm5,ccm12,ccm15,ccm20,ccm26,ccm27,ccm32,ccm35}
and so is not given here. The ground-state energy is given by 
\begin{equation}
E_g = \langle \Phi | e^{-S} H e^S | \Phi \rangle ~~ .
\end{equation}
This equation is a function of the ket-state correlation coefficients
only.

We differentiate between those  the model states that are coplanar 
and the single model state that is non-coplanar (or ``3D''). The analysis 
for the coplanar model state is given in Ref. \cite{farnell}
and we refer the interested reader to this publication for
more details. The non-coplanar model state IV has spins that make 
an angle $\theta$ to the plane perpendicular to the external field, 
as is also shown in Fig.~\ref{model_states}. 
We rotate the local spin axes 
of the spins such that all spins appear to point downwards for all 
four model states I, II, III, and IV in Fig.~\ref{model_states}.

We have four new Hamiltonians after rotation of the local spin axes 
of the spins such that all spins appear to point downwards for all 
four model states I, II, III, and IV in Fig.~\ref{model_states}(a-d). 

The Hamiltonian for model state I, Fig.~\ref{model_states} after 
rotation of the local spin axes is given by:
\begin{eqnarray}
H &=& 
\sum_{\langle i_A \rightarrow i_B \rangle} \biggl \{  
- \frac  14 (1+{\rm cos}(2\alpha)) ( s_{i_A}^+ s_{i_B}^+ + s_{i_A}^- s_{i_B}^- ) 
\nonumber \\ & & ~~~~~~~
+ \frac 14 (1-{\rm cos}(2\alpha)) ( s_{i_A}^+ s_{i_B}^- + s_{i_A}^- s_{i_B}^+ ) 
\nonumber \\ & & ~~~~~~~
-  {\rm cos}(2\alpha) s_{i_A}^z s_{i_B}^z 
+ \frac 12 {\rm sin}(2\alpha) ( s_{i_A}^z s_{i_B}^+ + s_{i_A}^z s_{i_B}^- )
 \nonumber \\ & & ~~~~~~~
- \frac 12 {\rm sin}(2\alpha) ( s_{i_A}^+ s_{i_B}^z + s_{i_A}^- s_{i_B}^z )
 \biggr \} \nonumber \\
    &+&    \sum_{\langle i_{B,C} \rightarrow i_{C,A} \rangle} \biggl \{  
- \frac  14 (1+{\rm sin}(\alpha)) ( s_{i_{B,C}}^+ s_{i_{C,A}}^+ + s_{i_{B,C}}^- s_{i_{C,A}}^- ) 
\nonumber \\ & & ~~~~~~~
+ \frac 14 (1-{\rm sin}(\alpha)) ( s_{i_{B,C}}^+ s_{i_{C,A}}^- + s_{i_{B,C}}^- s_{i_{C,A}}^+ ) 
\nonumber \\ & & ~~~~~~~
-  {\rm sin}(\alpha) s_{i_{B,C}}^z s_{i_{C,A}}^z + \frac 12 {\rm cos}(\alpha) ( s_{i_{B,C}}^z s_{i_{C,A}}^+ + s_{i_{B,C}}^z s_{i_{C,A}}^- )
\nonumber \\ & & ~~~~~~~
- \frac 12 {\rm cos}(\alpha) ( s_{i_{B,C}}^+ s_{i_{C,A}}^z + s_{i_{B,C}}^- s_{i_{C,A}}^z )
 \biggr \} \nonumber \\              
  & &
  \nonumber \\ &-& 
  \lambda \sum_{i_C}  s_{i_C}^z 
    + \lambda {\rm sin}(\alpha) \left(\sum_{i_A} s_{i_A}^z +  \sum_{i_B}    s_{i_B}^z\right) 
    \nonumber \\ &-& 
  \frac {\lambda}2 {\rm cos}(\alpha)  \sum_{i_A}  (s_{i_A}^+ + s_{i_A}^-)   
     + \frac {\lambda}2 {\rm cos}(\alpha)  \sum_{i_B}  (s_{i_B}^+ + s_{i_B}^-)\;,\label{rotH2}     
\end{eqnarray}
where the sum $\langle i_A \rightarrow i_B \rangle$ goes from sublattice $A$ to 
sublattice $B$ (and with directionality). Note that  $\langle i_{B,C} \rightarrow i_{C,A} 
\rangle$ indicates a sum that goes from sublattice $B$ to sublattice 
$C$ and sublattice $C$ to sublattice $A$, respectively (and with directionality). 

The Hamiltonian for model state III, Fig.~\ref{model_states} after 
rotation of the local spin axes is given by:
\begin{eqnarray}
H &=& 
 \sum_{\langle i_C \rightarrow i_{A,B} \rangle} \biggl \{  
 \frac  14 (-1+{\rm cos}(\alpha-\beta)) ( s_{{i_C}}^+ s_{{i_{A,B}}}^+ + s_{{i_C}}^- s_{{i_{A,B}}}^- ) \nonumber \\
& & ~~~~~~~
+ \frac 14 (1+{\rm cos}(\alpha-\beta)) ( s_{{i_C}}^+ s_{{i_{A,B}}}^- + s_{{i_C}}^- s_{{i_{A,B}}}^+ )
\nonumber \\ & & ~~~~~~~
    +  {\rm cos}(\alpha-\beta) s_{{i_C}}^z s_{{i_{A,B}}}^z 
\nonumber \\ & & ~~~~~~~
    + \frac 12 {\rm sin}(\alpha-\beta) ( s_{{i_C}}^+ s_{{i_{A,B}}}^z + s_{{i_C}}^- s_{{i_{A,B}}}^z )
\nonumber \\ & & ~~~~~~~
   - \frac 12 {\rm sin}(\alpha-\beta) ( s_{{i_C}}^z s_{{i_{A,B}}}^+ + s_{{i_C}}^z s_{{i_{A,B}}}^- ) \biggr \} 
\nonumber \\ &+& 
   \sum_{\langle i_A , i_B \rangle} \biggl \{  
               \frac 12 ( s_{i_A}^+ s_{i_B}^- + s_{i_A}^- s_{i_B}^+ ) + s_{i_A}^z s_{i_B}^z  \biggr \} 
      \nonumber \\
     &+& 
    \lambda {\rm sin}(\alpha)\left(\sum_{i_A} s_{i_A}^z +  \sum_{i_B}    s_{i_B}^z \right)
    + \lambda {\rm sin}(\beta) \sum_{i_C}  s_{i_C}^z 
      \nonumber \\
     &+& 
    \frac {\lambda}2 {\rm cos}(\alpha) \left\{ \sum_{i_A}  (s_{i_A}^+  + s_{i_A}^-)  
    + \sum_{i_B}  (s_{i_B}^+ + s_{i_B}^-) \right\} 
\nonumber \\ &+& 
    \frac {\lambda}2 {\rm cos}(\beta)  \sum_{i_C}  (s_{i_C}^+ + s_{i_C}^-) \; ,  
    \label{rotH3} 
\end{eqnarray}
where the sum $\langle i_C \rightarrow i_{A,B} \rangle$ goes from sublattice $C$ to 
sublattices $A$ and $B$ (with directionality) and $\langle i_A , i_B \rangle$ goes over 
each bond connecting the $A$ and  $B$ sublattices, but counting each one once only 
(and without directionality). We note that we have three sites in the unit cell for 
all of the models states used for the triangular lattice antiferromagnet. The Hamiltonian
for model state II, Fig.~\ref{model_states}, is a limiting case of Eqs. (\ref{rotH2})
and (\ref{rotH3}).

The Hamiltonian for the non-coplanar model state IV after 
rotation of the local spin axes is given by:
\begin{eqnarray}
H &=& 
\sum_{\langle i \rightarrow j \rangle} \biggl \{  
(\sin^2 (\theta) - \frac 12  \cos^2 (\theta)) s_{i}^z s_{j}^z
\nonumber \\ & & ~~~~~~~
+ \frac 14 \left( \frac 12 \sin^2 (\theta) - \cos^2 (\theta) -  \frac 12 \right) (s_{i}^+ s_{j}^+ + s_{i}^- s_{j}^-)
\nonumber \\ & & ~~~~~~~
+ \frac 14 \left(\cos^2 (\theta) -  \frac 12 \sin^2 (\theta) - \frac 12 \right) (s_{i}^+ s_{j}^- + s_{i}^- s_{j}^+)
\nonumber \\ & & ~~~~~~~
+ \frac {\sqrt{3}}4 {\rm cos}(\theta) \big(s_{i}^z \{ s_{j}^+ + s_{j}^- \} - \{ s_{i}^+ + s_{i}^- \} s_{j}^z\big)
 \biggr \} \nonumber \\ &+& 
 \lambda \sum_{i}  {\rm sin}(\theta)  s_{i}^z \nonumber \\ 
 &+&   
{\rm i}   \sum_{\langle i \rightarrow j \rangle} \biggl \{  
 \frac {\sqrt{3}}4  {\rm sin}(\theta) (s_{i}^+ s_{j}^- - s_{i}^- s_{j}^+)
\nonumber \\ & & ~~~~~~~
+ \frac 34 {\rm sin}(\theta)  {\rm cos}(\theta) \big(s_{i}^z \{ s_{j}^+ - s_{j}^- \} + \{ s_{i}^+ - s_{i}^- \} s_{j}^z\big)
 \biggr \} \nonumber \\
  & &
  \nonumber \\ &+& 
{\rm i}   \frac {\lambda}2 \sum_{i}  {\rm cos}(\theta) (s_{i}^+ - s_{j}^-) \;,
    \label{rotH4} 
\end{eqnarray}
where $\langle i \rightarrow j \rangle$ are those ``directional'' nearest-neighbor bonds 
on the triangular going from the $A$ sublattice to $B$ sublattice, $B$ sublattice to $C$ sublattice,
and $C$ sublattice to $A$ sublattice (in those directions only and not reversed).  
We see that this Hamiltonian now contains ``real and imaginary components''.
Henceforth, we shall take the expression ``real and imaginary components'' to mean 
that the rotated Hamiltonian contains explicit factors involving the imaginary number 
$i \equiv \sqrt(-1)$ and other explicit factors that do not involve this imaginary number. 
We note that $\theta$ is the angle that the spins make to the plane perpendicular
to the applied external field. Full details of the rotations used in the derivation
of this Hamiltonian are presented in Appendix B.

The manner in which CCM equations may be solved for the bra- and 
ket-state equations when correlation coefficients are allowed to be complex
is discussed in the Appendix, although we note that the problem essentially
reduces to a doubling of the number of CCM equations to be solved (i.e.,
for the real and imaginary components separately). However, we note 
that we use the LSUB$m$ approximation (in which clusters of $m$
contiguous sites limited to $m$ spin flips are included in $S$ and $\tilde S$) 
and that we consider the angles as free parameters in the CCM calculation. 
These angles are found by direct minimization 
of the CCM ground-state energy. This was achieved computationally at a given level of
LSUB$m$ approximation, and a minimum ground state energy with respect to these 
canting angles was also found computationally for a given fixed value of $\lambda$. 
We note that there was only one angle was needed for model states I and IV,
whereas two such angles were needed for model state III.  
Values of $\lambda$ were varied incrementally and the 
minimization process of the energy with respect to the canting angles repeated. 
The CCM calculations are costly in terms of computing time 
because we needed to minimize the ground-state energy with 
respect to such angles at each value of $\lambda$. We remark that the 
we have twice as many equations to solve for the 3D model state 
(for the real and imaginary components separately) ``as normal'' 
for the CCM. For this reason, results up to the LSUB6 level of approximation only are
quoted in these initial tests, although we find that ground-state energies
are highly converged for all model states even at this relatively low level 
of approximation.

In order to investigate the magnetization process in antiferromagnets, we consider  
the total lattice magnetization $M$ along the direction of the magnetic
field. We note that the important Hellmann-Feynman theorem is
obeyed by the CCM and so we may obtain the lattice magnetization
by finding $M = -\partial (E_g/N) / \partial \lambda$, which is carried out 
computationally in this paper. The method by which other expected values
may be found is also discussed in the Appendix. 

\section{Results}

As mentioned above, we present CCM results for model states I, II, III, 
and IV shown in Fig.~\ref{model_states}. The computational effort of the 
CCM calculations presented here 
for the non-coplanar model state IV to very high orders is very great and 
so results up to LSUB6 are presented here only for this model state 
in these initial studies. Results of the coplanar model states I to III from Ref.
\cite{farnell} up to LSUB6 are also presented here for the purposes of 
comparison only, although we note that higher orders of approximation than 
LSUB6 were carried out in Ref. \cite{farnell} for these states.

\begin{figure}
\epsfxsize=13cm
\centerline{\epsffile{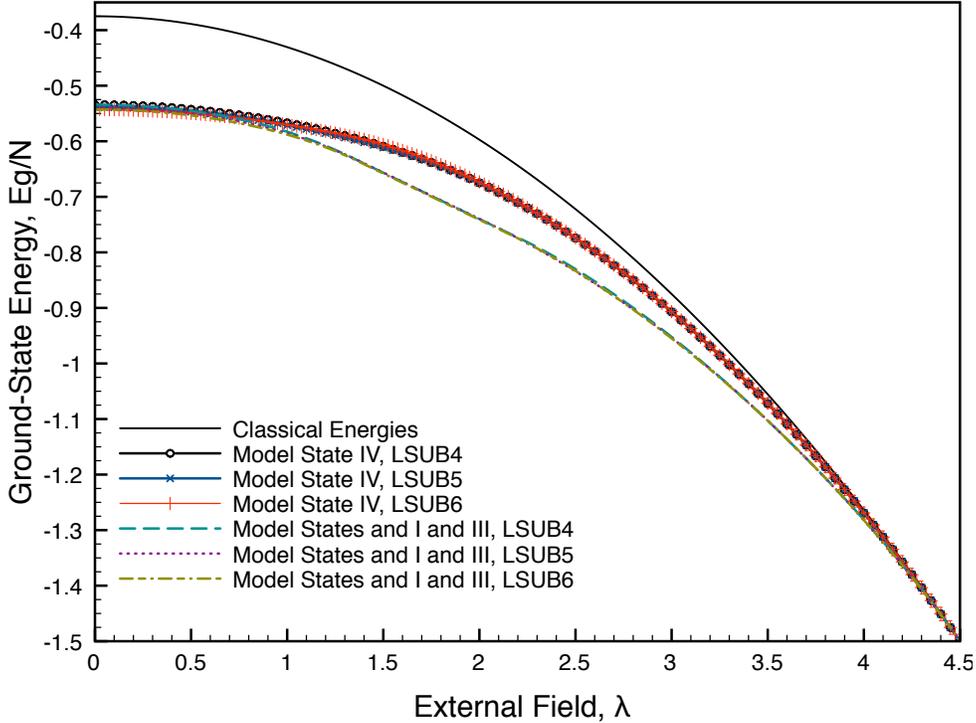}}
\caption{
The ground-state energy per site, $E_g/N$. Energies for the coplanar states I-III 
from Ref. \cite{farnell} are lower than those of the non-coplanar state IV for all $\lambda$.
}
\label{triangle_energies}
\end{figure}

The results for the ground-state energy are shown in Fig.~\ref{triangle_energies}.
We note that the results for the coplanar model states I to III from 
Ref. \cite{farnell} with lowest energy 
are shown only as a function of $\lambda$ in Fig.~\ref{triangle_energies}.
Thus, results of model state I only are presented for small values of
the applied magnetic field strength $\lambda$ and results of model state 
III only are presented for higher values of $\lambda$ near to 
$\lambda_s$. The results of both model states coincide in the intermediate
regime. Again, these LSUB$m$ series of results are found to converge rapidly with 
increasingly levels of LSUB$m$ approximation over all values of the 
external field parameter $\lambda$. 

Results for model state IV are also shown. Firstly, we note that the
imaginary component of the ground-state energy is found to sum 
to zero, as expected and required. Furthermore, the results for the ground-state 
energies for model state IV are much higher in value than their coplanar 
counterparts at identical levels of LSUB$m$ approximation. Note that 
LSUB6 was the highest level of approximation possible for this model
state in these initial tests and so we limit all presented results (for all model states) to this 
level of approximation in order to allow a direct and unbiased comparison.  However, 
we see that results for the ground-state energy are highly converged 
even at the LSUB6 level of approximation (by comparing results of LSUB6
to LSUB5 and LSUB4 in Fig.~\ref{triangle_energies}) and so these results are
adequate to establish that the ground-state energies are indeed lower for the coplanar
case.  As noticed previously \cite{kawamura,chub,zhito,cabra}, these results provide 
clear and strong evidence that the ground state of this system is coplanar.
This case has therefore been an excellent first test of new CCM based
on a non-coplanar ``three-dimensional'' model state.

\begin{figure}
\epsfxsize=13cm
\centerline{\epsffile{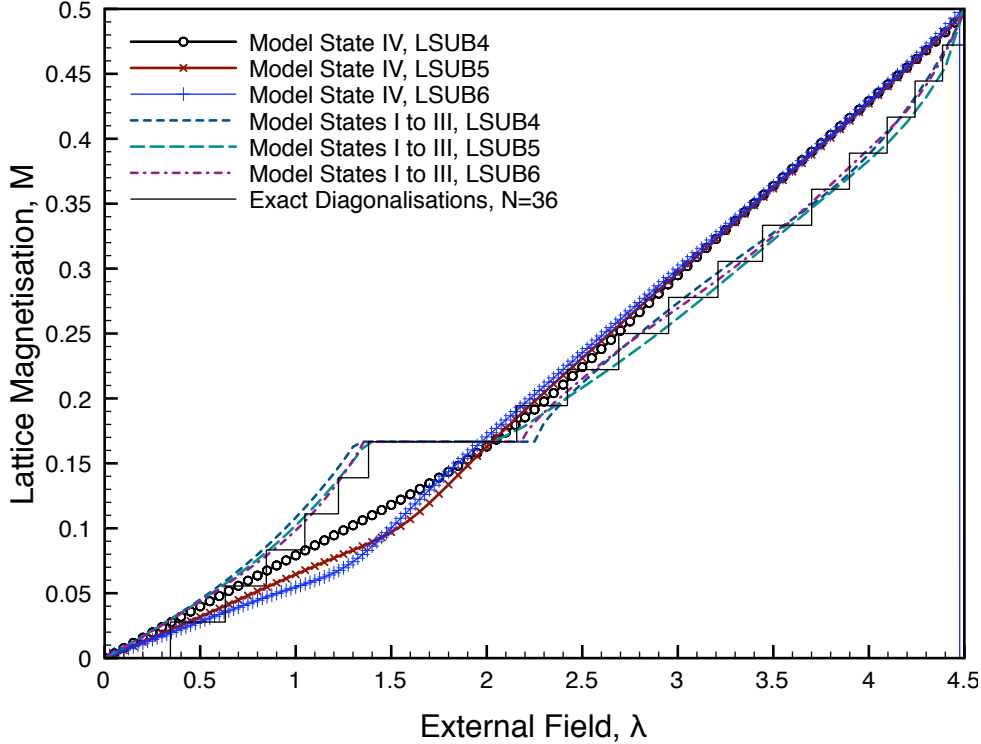}}
\caption{The lattice magnetization, $M$.
Results for model state IV do not detect the spin plateau at $M=1/6$.}
\label{triangle_magnetization}
\end{figure}

The results for the total lattice magnetization for model states I to III 
from Ref. \cite{farnell} are shown in Fig.~\ref{triangle_magnetization}. 
The LSUB$m$ results are again seen 
to converge rapidly for increasing $m$. However, there is a 
radical departure from the classical straight-line behavior (i.e.
$M_{\rm{Classical}}=\frac{1}{9}\lambda$) in this case. These 
previous CCM results for the coplanar states accurately detect  
the plateau in the $M$ versus $\lambda$ curve at $M=\frac 16$. 
Indeed, these previous results of the CCM \cite{farnell} for the 
coplanar model states carried out to the LSUB8 level of 
approximation indicate that the width of this 
plateau is given by $1.37 \lesssim \lambda \lesssim 2.15$.
(Note that the plateau corresponds to the ``straight" part of the curve in 
the $E_g(\lambda)$ curve shown in Fig.~\ref{triangle_energies}.)

By contrast, we note that the new results from model state IV show no such 
spin plateau any level of LSUB$m$ attempted (e.g., LSUB4, LSUB5 and LSUB6
shown in the Fig.~\ref{triangle_magnetization}). Note that this plateau has 
been observed for this system by other approximate methods and in experiment 
\cite{shirata}.  We note importantly again that the lattice magnetization must be 
real-valued because we find values for this quantity by taking first derivative of the 
ground-state energy, which itself is found to 
be real-valued (and not complex) for all values of $\lambda$. 
All of these results are excellent corroborating
evidence that the ground state is not of the type shown by model state IV (i.e.,
non-coplanar), but is rather of coplanar type shown by model states
I to III. Again, this is an excellent first check of the method for 
3D model states.

\section{Conclusions}

In this article we described how the coupled cluster method (CCM) 
may be applied to study the behavior of quantum magnets with the
aid of non-coplanar ``3D'' model states. A (slightly) modified method 
for solving for the CCM equation by ``direct iteration'' and for complex
CCM correlation coefficients is presented in an Appendix. We employed 
new computer code to find the ground-state energy and the total lattice 
magnetization of the spin-half triangular-lattice 
antiferromagnetic systems in the presence of external magnetic 
fields for both coplanar and non-coplanar model
states (which are the classical ground states). It was found that the
ground-state energy was real-valued for the non-coplanar model
state, as expected and required.
Results up to LSUB6 were possible only in these initial tests for the 
non-coplanar model state  due to the increased computational
complexity of the problem. However, these results were clearly
highly converged even at this level of approximation. In agreement 
with previous results of other methods, the coplanar
states were found to have lower energy.  Furthermore, the spin
plateau for $M=\frac 16$ known to occur in this system was detected 
by the coplanar model states, although it was not detected by the 
non-coplanar model state at any level of approximation. 
We conclude that the spin-half triangular lattice
antiferromagnet in an external magnetic field for the non-coplanar 
model state has been an excellent first test of new CCM based
for a non-coplanar ``three-dimensional'' model state.\\

\pagebreak 
\appendix
\section{Using 3D Model States For the CCM}

The method by which 3D model states may be employed for the high-order
CCM code is discussed below. The results for the application of this new
approach to the spin-half triangular lattice antiferromagnet in an external 
magnetic field are presented in the main text.

\subsection{Ket-State Equations}

Normally we solve the CCM equations either by using Newton-Raphson method
or by directly iterating the CCM equations. The direct iteration method
for the ket-state is achieved by casting the $N_f$ basic CCM equations
(each corresponding to one of the $N_f$ fundamental clusters at a given 
level of approximation), namely:
\begin{equation}
\langle\Phi|C_I^{-} {\rm e}^{-S} H {\rm e}^S|\Phi\rangle = f_I = 0 ~~ , 
\end{equation}
which we write as
\begin{equation}
 f_I({\rm consts.}, x_1, \cdots , x_{N_f} , x_1^2, \cdots , x_{N_f}^2 , x_1^3, \cdots , 
x_{N_f}^4 , x_1^4, \cdots , x_{N_f}^4 )=0  ~~ .
\end{equation}
We then rearrange this equation into the form 
\begin{equation}
\alpha_l \times x_l = f_I'({\rm consts.}, 
x_1, \cdots , x_{l-1}, x_{l+1}, \cdots, x_{N_f} , 
x_1^2, \cdots , x_{N_f}^2 , x_1^3, \cdots , 
x_{N_f}^4 , x_1^4, \cdots , x_{N_f}^4 ) ~~ .
\label{equation1}
\end{equation}
Hence, given an initial starting solution for the set of correlation coefficients
$\{ x_1, \cdots , x_{N_f} \}$, we iterate the following equation
for all of the $N_f$ equations  to convergence:
\begin{equation}
x_l = \frac {f_I'({\rm consts.}, 
x_1, \cdots , x_{l-1}, x_{l+1}, \cdots, x_{N_f} , 
x_1^2, \cdots , x_{N_f}^2 , x_1^3, \cdots , 
x_{N_f}^4 , x_1^4, \cdots , x_{N_f}^4 )} {\alpha_l }  ~~ .
\label{directItReal}
\end{equation}
This simple approach is successful in solving the CCM equations when all
of the correlation coefficients are real. 

However, for 3D model states we cannot ensure that the Hamiltonian is 
real-valued after rotation of the local spin axes. Indeed, in most cases it 
will not be so. The extension of the direct-iteration approach to complex
solutions for the CCM ket-state correlation coefficients 
might be achieved computationally by using complex data type (and all operators)
using {\tt C++} classes. However, the changes to the CCM code would be
extensive. A simpler alternative is just to break the $N_f$ CCM equations to 
$2 \times N_f$ equations, corresponding to the real and complex parts of the
equation(s) governed by Eq. (\ref{equation1}).

Hence, we split $x_l$ into real and imaginary components, i.e., 
$x_l = a_l + {\rm i} b_l$, as well as 
$\alpha_l =\beta_l + {\rm i} \gamma_l$
and  $f_l' = g_l  + {\rm i} h_l$ such that
\begin{eqnarray}
a_l& = & 
\frac { g_I({\rm consts.}, 
x_1, \cdots , x_{l-1}, x_{l+1}, \cdots, x_{N_f} , 
x_1^2, \cdots , x_{N_f}^2 , x_1^3, \cdots , 
x_{N_f}^4 , x_1^4, \cdots , x_{N_f}^4  ) + 
\gamma_l b_l }  {\beta_l} \nonumber \\
b_l & = & 
\frac { h_I({\rm consts.}, 
x_1, \cdots , x_{l-1}, x_{l+1}, \cdots, x_{N_f} , 
x_1^2, \cdots , x_{N_f}^2 , x_1^3, \cdots , 
x_{N_f}^4 , x_1^4, \cdots , x_{N_f}^4  ) - 
\alpha_l a_l}  {\gamma_l} \nonumber \\
&&\label{directItImag}
\end{eqnarray}
We now iterate these (now) $2 \times N_f$ equations for $a_l$ and $b_l$
(which correspond to the real and imaginary parts of ket-state correlation
coefficient $x_l$) to convergence. Note that we use the following 
products identities (up to fourth order) in the CCM equations of Eq. (\ref{directItImag}), 
namely:
\begin{eqnarray}
x_1 &=& a_1 + {\rm i} b_1  \nonumber \\
&&  \nonumber \\
x_1 \times x_2 &=& (a_1 + {\rm i} b_1) \times  (a_2 + {\rm i} b_2) = a_1 a_2 -  b_1 b_2 + {\rm i}(a_1 b_2 + b_1 a_2) \nonumber \\
&&  \nonumber \\
x_1 \times x_2 \times x_3 &=& 
x_1 \times x_2 \times  (a_3 + {\rm i} b_3)  \nonumber \\
&=& a_1 a_2 a_3 -  b_1 b_2 a_3  -  b_1 a_2 b_3  -  a_1 b_2 b_3   \nonumber \\
&&+ {\rm i}(b_1 a_2 a_3 + a_1 b_2 a_3+  a_1 a_2 b_3 -  b_1 b_2 b_3) \nonumber \\
&&  \nonumber \\
x_1 \times x_2 \times x_3 \times x_4 &=& 
x_1 \times x_2  \times x_3 \times  (a_4 + {\rm i} b_4)  \nonumber \\
&=& a_1 a_2 a_3 a_4 -  b_1 b_2 a_3  a_4 -  a_1 b_2 b_3 a_4 -  b_1 a_2 b_3  a_4   \nonumber \\
&& -  a_1 b_2 a_3 b_4 - b_1 a_2 a_3 b_4 - a_1 a_2 b_3 b_4  +  b_1 b_2 b_3 b_4 \nonumber \\
&&+ {\rm i}(  b_1 a_2 a_3 a_4 + a_1 b_2 a_3 a_4 +  a_1 a_2 b_3 a_4  + a_1 a_2 a_3 b_4  \nonumber \\ 
&&  -  a_1 b_2 b_3 b_4 -  b_1 a_2 b_3 b_4   -  b_1 b_2 a_3  b_4 -  b_1 b_2 b_3 a_4 ) \nonumber 
\end{eqnarray}
These identities have been coded directly into the high-order CCM code in order to solve for complex
ket-state coefficients; although this is somewhat tedious, it has the advantage of 
being more straightforward and probably a little speedier at runtime than using complex number classes
to solve this problem.

\subsection{Ground-State Energies}

Ground-state energies are found by evaluating
\begin{equation}
E_g = \langle \Phi | e^{-S} H e^S | \Phi \rangle ~~ .
\end{equation}
We note that the ground-state energy $E_g$ is a function of the CCM coefficients
up to quadratic terms in these coefficients 
(i.e., terms such as const., $x_1$ and $x_1 \times x_2$). Hence, 
we use the above equations in order to find the real and  imaginary parts of such 
terms. Furthermore, as the Hamiltonian has both real and imaginary parts we
divide the ground-state energies via:
\begin{equation}
E_g = E_g({\rm real}) + {\rm i} E_g({\rm imag.}) ~~.
\end{equation}
After determining the real and imaginary components of the CCM ket-state
correlation coefficient, we ought to find that $E_g({\rm imag.})=0$, as
all of the terms contributing to it should cancel each other out.  This is 
because that are real before rotation of local spin axes (and must
$E_g$ clearly is real number) must still be real even though we are
now solving for complex CCM correlation coefficients. 
Indeed, this constraint, i.e., $E_g({\rm imag.})=0$, 
forms an excellent check that our computer code is 
working correctly. This was found to be the case for model state 
IV employed here for the triangular-lattice Heisenberg antiferromagnet for 
all values of the external magnetic field strength, $\lambda$.

\subsection{Bra-State Equations}

We determine the 
bra-state correlation coefficient firstly by finding
\begin{equation}
\bar H = \langle \tilde \Psi | H | \Psi \rangle
\end{equation}
and then by taking the partial derivative of this expression with respect to 
$x_l$. We note again that the bra-state operator is defined via
$\tilde{S} =1 + \sum_{I \neq 0} \tilde{{\cal S}}_I C_I^{-}$ and so we
see that we obtain a linear equation 
\begin{equation}
\frac {\partial \bar H} {\partial x_l} =  \frac {\partial }{\partial x_l}  \langle \tilde \Phi | (1 + \sum_{I \neq 0} \tilde{{\cal S}}_I C_I^{-})
e^{-S} H e^S | \Phi \rangle = 0
\end{equation}
However, we note again that $E_g = \langle \Phi | e^{-S} H e^S | \Phi \rangle$ and
$\langle\Phi|C_I^{-} {\rm e}^{-S} H {\rm e}^S|\Phi\rangle = f_I$
and so we may write the above bra-state equation as
\begin{equation}
\frac {\partial \bar H} {\partial x_l} =     \frac {\partial E_g}{\partial x_l}   + \sum_{I \neq 0} \tilde{{\cal S}}_I 
\frac {\partial f_I }{\partial x_l} = 0
\end{equation}
This is just an equation that is linear in terms of the bra-state correlation
coefficients. If we now write $\tilde x_l = \tilde{{\cal S}}_I $ then  
this equation is solving readily by putting the $\tilde x_l $ terms 
for the $l^{\rm th}$ bra-state equation on the left-hand side of the equation
and everything else on the right-hand side. We then iterate to convergence
once again. Previously, bra-state correlation coefficients were real-valued,
although we cannot preclude that the bra-state coefficients are
complex for 3D model states. Once again we solve this problem 
by iterating $2 \times N_f$ bra-state equations, corresponding 
to the real and imaginary parts of the bra-state correlation coefficients. 
In this case, we have terms such as:

\begin{eqnarray}
\tilde x_1 &=& \tilde a_1 + {\rm i} \tilde b_1  \nonumber \\
&&  \nonumber \\
\tilde x_1 \times x_2 &=& (\tilde a_1 + {\rm i} \tilde b_1) \times  (a_2 + {\rm i} b_2) = 
\tilde a_1 a_2 -  \tilde b_1 b_2 + {\rm i}(\tilde a_1 b_2 + \tilde b_1 a_2) \nonumber \\
&&  \nonumber \\
\tilde x_1 \times x_2 \times x_3 &=& 
\tilde x_1 \times x_2 \times  (a_3 + {\rm i} b_3)  \nonumber \\
&=& \tilde a_1 a_2 a_3 -  \tilde b_1 b_2 a_3  -  \tilde b_1 a_2 b_3  -  \tilde a_1 b_2 b_3   \nonumber \\
&&+ {\rm i}(\tilde b_1 a_2 a_3 +\tilde  a_1 b_2 a_3+  \tilde a_1 a_2 b_3 -  \tilde b_1 b_2 b_3) \nonumber \\
&&  \nonumber \\
\tilde x_1 \times x_2 \times x_3 \times x_4 &=& 
\tilde x_1 \times x_2  \times x_3 \times  (a_4 + {\rm i} b_4)  \nonumber \\
&=& \tilde a_1 a_2 a_3 a_4 -  \tilde b_1 b_2 a_3  a_4 -  \tilde a_1 b_2 b_3 a_4 -  \tilde b_1 a_2 b_3  a_4   \nonumber \\
&& -  \tilde a_1 b_2 a_3 b_4 - \tilde b_1 a_2 a_3 b_4 - \tilde a_1 a_2 b_3 b_4  +  \tilde b_1 b_2 b_3 b_4 \nonumber \\
&&+\,\, {\rm i}( \tilde  b_1 a_2 a_3 a_4 + \tilde a_1 b_2 a_3 a_4 +  \tilde a_1 a_2 b_3 a_4  +\tilde  a_1 a_2 a_3 b_4  \nonumber \\ 
&&  -  \tilde a_1 b_2 b_3 b_4 - \tilde  b_1 a_2 b_3 b_4   -  \tilde b_1 b_2 a_3  b_4 -  \tilde b_1 b_2 b_3 a_4 ) \nonumber 
\end{eqnarray}

\subsection{Other Expectation Values}

As the well-known Helmann-Feynman theorem is obeyed by the CCM,
the lattice magnetization is found by taking the first derivative of the 
ground-state energy per spin with respect to the external field strength 
$\lambda$. This was achieved computationally in this article as this
approach is simple and straightforward. An alternative approach 
(not carried out here) may also be used to find the lattice magnetization 
$\langle s^z \rangle$. We begin by rotating the local coordinates of the 
spin axes and we form the relevant operator $\langle s^z \rangle$, 
which may well now be complex. However, once we have solved for 
both the ket-and bra-state equations we should find that the imaginary
part of this quantity should sum to zero, as was indeed seen for the 
ground-state energy.

We note that the local ``sublattice'' magnetization $\langle \tilde \Psi | s^z | \Psi \rangle$ 
(where $s^z$ is expressed in spin coordinates {\it after} rotation of
the local spin axes) is found to be complex. This seems perplexing at
first because we are finding $\langle \tilde \Psi | s^z | \Psi \rangle$, i.e.,
there appears to be no imaginary term inside this expectation value! However, 
we see that $s^z$ is expressed in terms of the {\it rotated} spin axes. If we were 
to carry out the {\it inverse} transformation (i.e., from the rotated spin axes to the 
original axes) then $s^z$ would, of course, probably become complex-valued 
when expressed in terms of these {\it original} spin axes. We might well now expect 
this new quantity to be complex-valued, and indeed this was found to be the 
case. Caution should therefore be exercised in attaching importance to the 
``sublattice magnetization'' for such 3D model states, although it might be
well-defined in certain circumstances. This is quite different to coplanar 2D model 
states for the CCM, and for which the sublattice magnetization is always 
real.

To summarize, however, only those macroscopic observables that 
are real-valued in the original spin coordinates for 3D model states ought 
definitely to be real-valued once we have solved for the ket- and bra-state
equations and formed the relevant expectation values.  In these cases, 
imaginary components present after rotation of local spin axes should 
sum to zero when evaluating the expectation value.

\section{Rotation of the local spin axes}

We now present a derivation of the Hamiltonian after rotation of the local
spin axes. We begin by again presenting the Hamiltonian before rotation of the 
spin axes:
\begin{equation}
H = \sum_{\langle i,j\rangle} {\bf s}_i ~ \cdot ~ {\bf s}_j - \lambda \sum_i s_i^y
~~ ,
\label{newH}
\end{equation}
where the index $i$ runs over all lattice sites on the triangular lattice. The
expression $\langle i,j\rangle$ indicates a sum over all nearest-neighbor 
pairs, although each pair is counted once and once only. The strength 
of the applied external magnetic field is given by $\lambda$.
Note that we assume that the external field applies now in the $y$-direction
rather than the $z$-direction. This is a notational change only 
in order to simplify somewhat the following mathematics and it does
not affect our results. Indeed, this choice allows us to use the model state with 
zero field now lies in the $xz$-plane, as used in Ref. \cite{ccm12}.

The classical ground-state of Eq. (\ref{newH}) when $\lambda=0$ is 
the N\'eel-like state where all spins on each sublattice are separately aligned 
(again as in Ref. \cite{ccm12}, all 
in the $xz$-plane, say). The spins on sublattice A 
are oriented along the negative {\em z}-axis, and spins on sublattices 
B and C are oriented at $+120^\circ$ and $-120^\circ$, respectively, 
with respect to the spins on sublattice A. We perform 
the following spin-rotation transformations. 
Specifically, we leave the spin axes on sublattice A unchanged, and we 
rotate about the $y$-axis the spin axes on sublattices B and C by 
$-120^\circ$ and $+120^\circ$ respectively, 
\begin{eqnarray} 
s_B^x \rightarrow -\frac{1}{2} s_B^x - \frac{\sqrt{3}}{2} s_B^z \;\;  &;& \;\; 
s_C^x \rightarrow -\frac{1}{2} s_C^x + \frac{\sqrt{3}}{2} s_C^z \;\; , 
\nonumber \\
s_B^y \rightarrow s_B^y \;\; &;& \;\; s_C^y \rightarrow s_C^y \;\; , 
\nonumber \\ 
s_B^z \rightarrow  \frac{\sqrt{3}}{2} s_B^x -\frac{1}{2} s_B^z \;\; &;& \;\; 
s_C^z \rightarrow -\frac{\sqrt{3}}{2} s_C^x -\frac{1}{2} s_C^z \;\; . 
\label{eq20}
\end{eqnarray} 
This leads to a Hamiltonian given by:
\begin{equation} 
H = \sum_{\langle i\rightarrow j\rangle}
\left \{
-{1\over 2} s_i^x s_j^x  + s_i^y s_j^y -{1\over 2} s_i^z s_j^z 
+\frac{\sqrt{3}}{2}
( s_i^z s_j^x -s_i^x s_j^z )\right \} - \lambda \sum_i s_i^y
\label{Hstage1}
\end{equation}
We note that the summation in Eq. (\ref{Hstage1}) 
again runs over nearest-neighbor
bonds, but now also with a {\it directionality} indicated by 
$\langle i \rightarrow j\rangle$, which goes from A to B, B to C, and 
C to A. Spins now appears mathematically to point in the downwards
$z$-direction when the external field is zero ($\lambda=0$). For
$\lambda > 0$, the spins appear now to be at an angle of $\theta$ to 
the negative $z$-axis in the $yz$-plane. Thus, we rotate the local 
spin axes additionally by an amount $+\theta$ about the $x$-axis by 
using the following (passive) transformation: 
\begin{equation} 
s^x \rightarrow s^x ~~;~~ s^y \rightarrow s^y \cos(\theta) - s^z \sin(\theta) 
~~ ; ~~ s^z \rightarrow s^z \cos(\theta) + s^y \sin(\theta) ~~ .
\end{equation}
This transformation yields a Hamiltonian of:
\begin{eqnarray}
H &=&  \sum_{\langle i\rightarrow j\rangle} 
\Bigl \{
-{1\over 2} s_i^x s_j^x  + (\cos^2(\theta) - \frac 12 \sin^2(\theta)) s_i^y s_j^y 
 \nonumber \\
&& ( \sin^2(\theta) - \frac 12 \cos^2(\theta) ) s_i^z s_j^z 
-\frac{3}{2} \cos(\theta) \sin(\theta) ( s_i^y s_j^z + s_i^z s_j^y )
 \nonumber \\
&& 
+\frac{\sqrt{3}}{2} \cos(\theta) ( s_i^z s_j^x -s_i^x s_j^z )
+\frac{\sqrt{3}}{2} \sin(\theta) ( s_i^y s_j^x -s_i^x s_j^y )
\Bigr \} \nonumber \\
&&
- \lambda \sum_i \{ s_i^y \cos(\theta) - s_i^z \sin(\theta) \}
\label{Hstage2}
\end{eqnarray}
All spins now appear to the lie in the negative $z$-direction 
for any given value thus of $\theta$  (and thus also for all values 
of the external field). We now use the following expressions:
\begin{eqnarray}
&&s^{\pm} = s^x \pm {\rm i} s^y ~~;~~ s^x = \frac 12 (s^+ + s^ -)  ~~ ;~~ s^y=\frac {\rm i}2 (s^- - s^ +)  
 \nonumber \\
 &&s^x s^x = \frac 14 (s^+ s^+ +s^+ s^- + s^- s^+ + s^- s^-)  ~~ ;~~ 
 s^y s^y = \frac 14 (-s^+ s^+ +s^+ s^- + s^- s^+ - s^- s^-)  
  \nonumber \\
 &&s^x s^y = \frac {\rm i}4 (-s^+ s^+ - s^- s^+ + s^+ s^- + s^- s^-)  ~~ ;~~ 
 s^y s^x = \frac {\rm i}4 (-s^+ s^+ - s^+ s^- + s^- s^+ + s^- s^-)  
   \nonumber
 \end{eqnarray}
such that the final version of the Hamiltonian is obtained:
\begin{eqnarray}
H &=& 
\sum_{\langle i \rightarrow j \rangle} \biggl \{  
(\sin^2 (\theta) - \frac 12  \cos^2 (\theta)) s_{i}^z s_{j}^z
\nonumber \\ & & ~~~~~~~
+ \frac 14 \left( \frac 12 \sin^2 (\theta) - \cos^2 (\theta) -  \frac 12 \right) (s_{i}^+ s_{j}^+ + s_{i}^- s_{j}^-)
\nonumber \\ & & ~~~~~~~
+ \frac 14 \left(\cos^2 (\theta) -  \frac 12 \sin^2 (\theta) - \frac 12 \right) (s_{i}^+ s_{j}^- + s_{i}^- s_{j}^+)
\nonumber \\ & & ~~~~~~~
+ \frac {\sqrt{3}}4 {\rm cos}(\theta) \big(s_{i}^z \{ s_{j}^+ + s_{j}^- \} - \{ s_{i}^+ + s_{i}^- \} s_{j}^z\big)
 \biggr \} \nonumber \\ &+& 
 \lambda \sum_{i}  {\rm sin}(\theta)  s_{i}^z \nonumber \\ 
 &+&   
{\rm i}   \sum_{\langle i \rightarrow j \rangle} \biggl \{  
 \frac {\sqrt{3}}4  {\rm sin}(\theta) (s_{i}^+ s_{j}^- - s_{i}^- s_{j}^+)
\nonumber \\ & & ~~~~~~~
+ \frac 34 {\rm sin}(\theta)  {\rm cos}(\theta) \big(s_{i}^z \{ s_{j}^+ - s_{j}^- \} + \{ s_{i}^+ - s_{i}^- \} s_{j}^z\big)
 \biggr \} \nonumber \\
  & &
  \nonumber \\ &+& 
{\rm i}   \frac {\lambda}2 \sum_{i}  {\rm cos}(\theta) (s_{i}^+ - s_{j}^-) \;,
 \label{finalH}
\end{eqnarray}
The model state $|\Phi \rangle$ now consists of spins that appear 
(mathematically only) to lie along the negative $z$-axis for all values 
of $\theta$. We note that by using $E = \langle \Phi | H | \Phi \rangle$ and also 
by using the expressions
\begin{equation}
\langle \Phi | s^z|\Phi\rangle = -\frac 12
~~ ; ~~ \langle \Phi | s^z s^z |\Phi\rangle =  \frac 14
~~ ; ~~ \langle \Phi | s^\pm |\Phi\rangle = 
~ \langle \Phi | s^\pm s^\pm |\Phi\rangle ~ = 0  ~~ ,
\end{equation}
we obtain the classical ground-state energy by using Eq. (\ref{finalH}), 
namely, of $E/N = (3/4) \{ \sin^2(\theta)-0.5 \cos^2(\theta) \} - (\lambda / 2) \sin(\theta)$, 
as required. This is an excellent initial check of the the Hamiltonian of Eq. (\ref{finalH}) 
after all of the rotations of the local spin axes have been carried out.

\pagebreak

\end{document}